\documentclass[a4paper,10pt,twoside]{cpc-hepnp}

\usepackage{multicol}
\usepackage{graphicx}
\usepackage{booktabs}
\usepackage{amssymb,bm,mathrsfs,bbm,amscd}
\usepackage[tbtags]{amsmath}
\usepackage{lastpage}
\usepackage{CJKutf8}

\begin{document}
\begin{CJK*}{UTF8}{gbsn}

\fancyhead[co]{\footnotesize LU Meng-Jiao et al: Diffusion monte carlo calculations of three-body systems}

\footnotetext[0]{Received 14 February 2012}

\title{Diffusion monte carlo calculations of three-body systems\thanks{
Supported by National Natural Science Foundation
of China (10735010, 10975072, 11035001, 11120101005),
973 National Major State Basic Research and Development of China
(2007CB815004 ,2010CB327803), CAS Knowledge
Innovation Project (KJCX2-SW-N02), Research Fund of Doctoral
Point (20100091110028), Project Funded by the
Priority Academic Program Development of Jiangsu Higher Education
Institutions (PAPD).
}}

\author{%
      LU Meng-Jiao(吕梦蛟)$^{1;1)}$\email{mengjiao.lu@gmail.com}%
\quad REN Zhong-Zhou(任中洲)$^{1,2}$
\quad Lin Qi-Hu(林祁斛)$^{1}$
}
\maketitle

\address{%
$^1$ Department of Physics, Nanjing University, Nanjing 210093, China\\
$^2$ Center of Theoretical Nuclear Physics, National Laboratory of
Heavy Ion Accelerator, Lanzhou 730000, China\\
}

\begin{abstract}
Application of diffusion Monte Carlo algorithm in three-body systems
is studied. We develop a program and use it to calculate the property
of various three-body systems. Regular Coulomb systems such as atoms,
molecules and ions are investigated. Calculation is then extended to
exotic systems where electrons are replaced by muons. Some nuclei with
neutron halos are also calculated as three-body systems consisting of
a core and two external nucleons. Our results agree well with
experiments and others' work. 


\end{abstract}



\begin{multicols}{2}

\section{Introduction}

The three-body problem plays an important role in atomic physics,
molecular physics and nuclear physics. 
The study of the Coulomb three-body systems in quantum physics dates
back to the 1930s when physicists were trying to explain the helium
spectrum. Many years have passed, but progresses are still being made
in this field. New systems such as exotic three-body atomic systems
have aroused great interest of phycists recently \cite{anacarani}. New
wave functions are also coming out \cite{simple}.
The study of three-body problem is also very common in nuclear physics
\cite{zhukov,effimov}. Tritium is a typical three-body system that has
been studied by Monte Carlo methods long ago \cite{tritium}. Some
exotic nuclei such as ${}^{6}$He and ${}^{11}$Li are also studied from
a three-body perspective
\cite{jensen,tostevin,thompson-6He,ren-JPG,ren-PRC,chu-halo}.

The fundamental difficulty in these fields is to solve the coupled
three-body Schrodinger equation. Variational method with trial wave
function is one of the most common ways to solve this problem
\cite{suzuki}. Some of these variational wave functions could be
analytical and  simple \cite{simple} while some of them are in
sophisticated forms with a large number of variational parameters
\cite{frolov-three-body}. Another way is to solve the corresponding
Faddeev equations which is used widely both in scattering and bound
problems \cite{faddeev,thompson}.
The diffusion Monte Carlo method is also very powerful in solving
three-body problems such as Positronium-atom complexes
\cite{atom-complexes}, mesic molecules \cite{mesic-molecule}, light
nuclei \cite{light-nuclei,annual} and medium nuclei
\cite{medium-nuclei}.

We write a program to study three-body systems with diffusion Monte
Carlo methods. We first study regular Coulomb systems such as hydrogen
molecular ion $\mathrm{p}^+\mathrm{p}^+\mathrm{e}^-$, followed by the
extension to muonic systems. Those calculations are carried out
without Born-Oppenheimer approximation. We also apply diffusion Monte
Carlo method to nuclei which are considered as Yukawa three-body
systems.


\section{Theoretical framework}
The Hamiltonian of a three-body system can be written as:
\begin{equation}\label{hamiltonian}
	H =-\sum_{i}{\frac{\hbar^2}{2m_i} \nabla_i^2 } + \sum_{i<j}{V_{ij}} \ ,
\end{equation}
where $m_i$ is the mass of each component and $V_{ij}$ is the two-body
potential.
The exact wave function of the three-body system is written as
$\psi({\bf R},t)$, where ${\bf R}$ is the 3N-dimension configuration
coordinate. Then a new wave function
\begin{equation}
	f({\bf R},t)=\psi({\bf R},t) \psi_T({\bf R},t)
\end{equation}
is introduced, where $\psi_T({\bf R},t)$ is a trial wave function
which could be generated by variational Monte Carlo method. This
function satisfies the equation \cite{umrigar-dmc-algorithm},
\begin{eqnarray}\label{basic_equation}
	-\frac{\partial}{\partial t} f({\bf R},t) &=&
\sum_i{-\frac{\hbar^2}{2m_i}\nabla_i^2 f({\bf R},t)}	\\ \nonumber &&+
\sum_i{\frac{\hbar^2 }{m_i} \nabla \cdot \left[\frac{\nabla_i
\psi_T({\bf R})}{\psi_T({\bf R})} f({\bf R},t)\right]}	\\ \nonumber
&&+ \left[E_L({\bf R})-E_T(t)\right] f({\bf R},t) \ . 
\end{eqnarray}
Here $E_L$ is the local energy given by
\begin{equation}
	E_L({\bf R})=\frac{{\hat H}_L \psi_T({\bf R})}{\psi_T({\bf R})} \ ,
\end{equation}
where
\begin{equation}
	{\hat H}_L =-\sum_i \frac{\hbar^2}{2m_i}\nabla_{i}^2 + \sum_{i<j}{V_{ij}} 		\ .
\end{equation}
$E_T(t)$ is a shift energy which plays a role of normalization factor.
Equation \ref{basic_equation} could be solved iteratively in an
integral form,
\begin{equation}
	f({\bf R}',t+\tau)=\int{d{\bf R} \tilde G({\bf R}',{\bf R},\tau) f({\bf R},t)} \ ,
\end{equation}
where $\tau$ is the time step between iterations. In short-time
approximation the Green's function $\tilde G({\bf R}',{\bf R},\tau)$
has the form \cite{umrigar-dmc-algorithm},
\begin{eqnarray}\label{short-time-green's-function}
	\tilde G({\bf R}',{\bf R},\tau) &=& \prod_i
		\frac{1}{(2\pi\sigma_i^2)^{3/2}}
		\exp\left[-\frac{({\bf R}'-\mu_i({\bf
	R}))^2}{2\sigma_i^2}\right] \\ \nonumber &&	\times
	\exp[-\tau(E_L({\bf R}')+E_L({\bf R})-2 E_T)/2] \ , 
\end{eqnarray}
where 
\begin{equation}
	\sigma_i^2=\tau\hbar^2/m_i \ ,
\end{equation}
and
\begin{equation}
	\mu_i({\bf R})={\bf R}+\sigma_i^2
		\nabla_i\ln|\psi_T({\bf R})| \ . 
\end{equation}
To simulate this short-time Green's function in calculation, initial
walkers will be generated according to the trial function $\psi_T$
first. Then the walkers will diffuse and drift to new positions
according to the Gaussian distribution
\begin{equation}
	\frac{1}{(2\pi\sigma_i^2)^{3/2}}
	\exp\left[-\frac{({\bf R}'-\mu_i({\bf R}))^2}{2\sigma_i^2}\right] \ .
\end{equation}
Then a branching technique is adopted to give the probability to kill
or multiply the walkers at new configurations. The number of copies
generated from each old walker is ${\text{INT}}(p+\xi)$, where $\xi$
is a random number between (0,1), and,
\begin{equation}
	p=\exp[-\tau(E_L({\bf R}')+E_L({\bf R})-2 E_T)/2] \ .
\end{equation}
Iteration times are denoted by ${\hat t}=t/\tau$. At each iteration,
the expectation value of energy is calculated by the mixed estimator,
which is defined as \cite{umrigar-dmc-algorithm},
\begin{equation}
	E_{\mathrm{mix}}(T)=
	\frac{
		\sum_{{\hat t}=0}^{T}\Pi({\hat t})
			\sum_{\alpha=1}^{N({\hat t})}
		\frac{{\hat H}_L \psi_T({\bf R}_\alpha({\hat t}))}{\psi_T({\bf R}_\alpha({\hat t}))}
		}{
		\sum_{{\hat t}=0}^{T}\Pi({\hat t})N({\hat t})
	} \ ,
\end{equation}
where 
\begin{equation}
\Pi({\hat t})=\prod_{m=0}^{{\hat t}}e^{-\tau E_T({\hat t}-m)} \ .
\end{equation}
To make the population of walkers stable, the $E_T$ should be adjusted
by
\begin{equation}
	E_T({\hat t}+1)=E_{\mathrm{mix}}({\hat t})-\log\frac{N({\hat t})}{N_0} \ ,
\end{equation}
where $N({\hat t})$ is the population at each iteration and $N_0$ is
the initial one.

After many iterations, these walkers will distribute as $\psi_T({\bf
R})\psi_0({\bf R})$, where $\psi_0$ is the exact ground state of the
system. $E_{\mathrm{mix}}(t)$ will also equal the exact ground state
energy. To avoid time-step error, the results of different $\tau$
should be calculated and then extrapolated to $\tau=0$.

\section{Numerical results and analyses}
We calculate different systems of atoms, molecules, ions and nuclei
with diffusion Monte Carlo method. In these calculations except
nuclei, the trial wave functions have variational form:
\begin{equation}\label{trial_function}
	\psi_T=\prod_{i<j}{e^{ \alpha_l r_{ij} }}.
\end{equation}
For nuclei, the trial wave functions are taken as:
\begin{equation}
	\psi_T=\prod_{i<j}{e^{ \alpha_l r_{ij} }}+\gamma \prod_{i<j}{e^{ \beta_l r_{ij} }}.
\end{equation}
The parameters $\{\alpha_l\}$, $\{\beta_l\}$ and $\gamma$ are
optimized by a variational Monte Carlo program. The number of initial
walkers is taken to be 2000. Each system is calculated with eight
different $\tau$. For each $\tau$, the iteration continues until the
error bar of energy, taken from the average of last 10000 mixed
estimators, is smaller than expected. The final result is then
constructed by extrapolation from the results of different $\tau$ to
$\tau=0$. 


\subsection{Atoms, molecules and ions}

We make diffusion Monte Carlo calculations of some regular Coulomb
three-body systems such as atoms, molecules and ions. Our calculations
are performed without Born-Oppenheimer approximation.
The introduction of nuclear degrees of freedom will increase running
time, but it is acceptable for few-body systems.

The calculation results are listed in Table \ref{atom}. The results
from experiments or accurate variational calculations are also listed
for comparison. For systems with only one heavy core, such as helium
atom, variational calculation results are already very close to the
experimental values. But for systems with two heavy cores, such as
hydrogen molecular ion $\mathrm{p}^+\mathrm{p}^+\mathrm{e}^-$,
variational results are much higher than the experimental value. This
is not surprising because we only use a quite simple trial wave
function. However, this simple wave function is good enough to be an
input of diffusion Monte Carlo program.

\begin{center}
\tabcaption{Results of regular Coulomb systems. $E_{\mathrm{VMC}}$ is
		the variational Monte Carlo result of ground state energy from
		optimizing the trial wave function. $E_{\mathrm{DMC}}$ is the
		ground state energy result from diffusion Monte Carlo
		calculation. The compared results from others' work or
		experiments are also listed. All values are in atomic
		units.\label{atom}}
\footnotesize
	\begin{tabular}{llllll}
	\hline\hline
		System			&$E_{\mathrm{VMC}}$	&$E_{\mathrm{DMC}}$
		&\multicolumn{2}{l}{Results in Refs.}\\
	\hline
		$\mathrm{e}^+\mathrm{e}^-\mathrm{e}^-$		&-0.2372	&-0.2614
		&-0.2620	&Frolov \cite{frolov-positronium-ion}\\
		$\mathrm{p}^+\mathrm{e}^-\mathrm{e}^-$		&-0.5061	&-0.5273
		&-0.5274	&Frolov \cite{frolov-ions}\\
		$\mathrm{p}^+\mathrm{p}^+\mathrm{e}^-$		&-0.4759	&-0.5938
		&-0.5974 	&Exp. \cite{zhang-H2+}\\
		${}^4$He$^{2+}\mathrm{e}^-\mathrm{e}^-$		&-2.886		&-2.902
		&-2.902 	&Exp. \cite{griffiths} \\
		${}^7$Li$^{3+}\mathrm{e}^-\mathrm{e}^-$		&-7.253		&-7.275
		&-7.279		&Ancarani \cite{anacarani}\\
	\hline\hline
	\end{tabular}
\end{center}

Most of these diffusion Monte Carlo results agree with experiment very
well with an error of only 0.05\%. This proves that our program is
accurate enough to calculate three-body system combined by Coulomb
interaction. The result of hydrogen molecular ion has a larger error
about 0.6\%. This larger error may be due to the fact that it has two
heavy cores and the trial wave function with the form of Eq.
\ref{trial_function} is not a good description of the system. 

\subsection{Muonic systems}
We also calculate some muonic three-body systems.  Unlike regular
three-body systems, the Born-Oppenheimer approximation will cause
serious error when the electrons are replaced by muons. In these
systems, the motion of nuclei cannot be omitted, since the mass of
muon is comparable to proton or light nuclei. Therefore our diffusion
Monte Carlo calculations of muonic systems are performed without
Born-Oppenheimer approximation. The calculation results are listed in
Table \ref{hyperatom}. Results for comparison are also listed. 

\begin{center}
\tabcaption{Results of muonic systems. $E_{\mathrm{VMC}}$ is the
		variational Monte Carlo result of ground state energy from
		optimizing the trial wave function. $E_{\mathrm{DMC}}$ is the
		ground state energy result from diffusion Monte Carlo
		calculation. The compared results from others' work or
		experiments are also listed. All values are in atomic
		units.\label{hyperatom}}
\footnotesize
	\begin{tabular}{llllll}
	\hline\hline
		System						&$E_{\mathrm{VMC}}$	&$E_{\mathrm{DMC}}$
		&\multicolumn{2}{l}{Results in Refs.}\\
	\hline
 		$\mathrm{\mu}^+\mathrm{e}^-\mathrm{e}^-$		&-0.5023	&-0.5228
 		&-0.5251 	&Frolov \cite{frolov-ions} \\
		$\mathrm{\mu}^+\mathrm{\mu}^+\mathrm{e}^-$		&-0.4685	&-0.5832
		& 		& \\
		$\mathrm{\mu}^+\mathrm{\mu}^-\mathrm{\mu}^-$		&-49.21		&-54.07
		& 		& \\
		$\mathrm{p}^+\mathrm{\mu}^-\mathrm{\mu}^-$		&-92.71		&-96.97
		&-97.57		&Frolov \cite{frolov-1993}\\
		$\mathrm{p}^+\mathrm{p}^+\mathrm{\mu}^-$		&-87.78		&-101.8
		&-96.86 	&Bailey\cite{frolov-universal}\\
		He$^{2+}\mathrm{\mu}^-\mathrm{\mu}^-$			&-579.1		&-582.3
		&-582.4		&Ancarani \cite{anacarani}\\
	\hline\hline
	\end{tabular}
\end{center}

Some of muonic systems in our calculations are rarely studied before
such as $\mathrm{\mu}^+\mathrm{\mu}^+\mathrm{e}^-$. So only diffusion
Monte Carlo results are given. Some of these systems have been
calculated with accurate variational methods in others' work. Our
results of these systems agree very well with them. Bigger difference
can be found in the calculation of system
$\mathrm{p}^+\mathrm{p}^+\mathrm{\mu}^-$, and our result is lower than
the result in Ref. \cite{frolov-universal}. Considering that our
diffusion Monte Carlo results are always a little bit higher than the
accurate values, this difference is very strange and should be
confirmed by more theoretical work or experiments.  

\subsection{Nuclei}
In three-body model, a nucleus can be treated as a system composed of
a core and two external nucleons. \cite{ren-xu} This is particularly
useful in the study of exotic nuclei with neutron halos. Three typical
halo nuclei, ${}^{11}$Li, ${}^{14}$Be and ${}^{17}$B, were studied by
equivalent two-body methods and Faddeev equations before
\cite{ren-JPG,ren-PRC,chu-halo}. We investigate these nuclei with
diffusion Monte Carlo method. The two-body Yukawa potential is taken
from Ref. \cite{ren-PRC}. The calculation results and various compared
values are listed in Table \ref{nuclei}. 

\begin{center}
\tabcaption{Results of some exotic nuclei\label{nuclei}. E is the
		ground state energy. $R_m$ is the matter root-mean-square
		radius. The lines denoted by (F\&R) are the results from
		equivalent two-body methods \cite{ren-JPG}. The lines denoted
		by (Faddeev) are the results from Faddeev equations
		\cite{chu-halo}. }
\footnotesize
	\begin{tabular}{llllll}
	\hline\hline
		System		&		&E			&$R_m$	\\
		  		&		&(MeV)			&(fm) \\
	\hline
		${}^{11}$Li	&(DMC)		&-0.59			&2.83		\\
				&(Faddeev)	&-0.54			&2.95		\\
				&(F\&R)		&-0.35 			&3.18		\\
				&(exp.)		&-0.35($\pm$0.05)	&3.10($\pm$0.17)\\
		${}^{14}$Be	&(DMC)		&-1.18			&2.78		\\
				&(Faddeev)	&-1.07			&2.85		\\
				&(F\&R)		&-1.12 			&2.90		\\
				&(exp.)		&-1.12($\pm$0.20)	&3.10($\pm$0.30)\\
		${}^{17}$B	&(DMC)		&-1.09			&2.73		\\
				&(Faddeev)	&-1.01			&2.76		\\
				&(F\&R)		&-0.84 			&2.81		\\
				&(exp.)		&-1.49($\pm$0.20)	&3.00($\pm$0.40)\\
	\hline\hline
	\end{tabular}
\end{center}

Our results of ground state energy are lower than the equivalent
two-body methods, but agree with the Faddeev equation well.
Considering that the equivalent two-body methods are variational
methods, our results and Faddeev results are better. These results
show that the diffusion Monte Carlo method can be as precise as
Faddeev equations in the calculations of three-body systems. However,
the diffusion Monte Carlo method can be also used to calculate
many-body systems, which is difficult for Faddeev equations.


\section{Summary}
In this paper, diffusion Monte Carlo algorithm with importance
sampling technique is formulated for systems consisting of components
with different masses. A mixed estimator is used to obtain the average
of physical quantities. We write a program and study various systems
with this method. We calculate the ground state energy of regular and
exotic three-body systems.  These calculations are all performed
without Born-Oppenheimer approximation. Our results agree very well
with experiments and with other high precision variational methods. We
also produce the properties of some three-body systems which are
rarely studied before. Halo nuclei are investigated as three-body
systems. The results are better than equivalent two-body method but
almost the same as that from the Faddeev equations. All these results
have proven the accuracy of our program and the power of diffusion
Monte Carlo method in studying the three-body systems.

\end{multicols}

\vspace{-1mm}
\centerline{\rule{80mm}{0.1pt}}
\vspace{2mm}

\begin{multicols}{2}

\end{multicols}

\clearpage

\end{CJK*}
\end{document}